\documentstyle[pra,aps,amsfonts,multicol,graphicx]{revtex}
\newcommand{\be}[1]{\begin{equation} #1 \end{equation}}
\newcommand{\bea}[1]{\begin{eqnarray} #1 \end{eqnarray} }
\newcommand{\ba}[2]{\left(\begin{array}{#1}#2\end{array}\right)}

\newcommand{\tr}[1]{{\rm Tr}\left(#1\right)}

\draft
\title{On the geometry of entangled states.}
\author{Frank Verstraete, Jeroen Dehaene, Bart De Moor}
\address{Katholieke Universiteit Leuven,
Department of Electrical Engineering, Research Group SISTA\\
Kard. Mercierlaan 94, B-3001 Leuven, Belgium }
\begin{document}

\pagestyle{plain} \pagenumbering{arabic}

\maketitle
\begin{abstract}
The basic question that is addressed in this paper is  finding
the closest separable state for a given entangled state, measured
with the Hilbert Schmidt distance. While this problem is in
general very hard, we show that the following strongly related
problem can be solved: find the Hilbert Schmidt distance of an
entangled state to the set of all partially transposed states. We
prove that this latter distance can be expressed as a function of
the negative eigenvalues of the partial transpose of the
entangled state, and show how it is related to the distance of a
state to the set of positive partially transposed states
(PPT-states). We illustrate this by calculating the closest
biseparable state to the W-state, and give a simple and very
general proof for the fact that the set of W-type states is not
of measure zero. Next we show that all surfaces with states whose
partial transposes have constant minimal negative eigenvalue are
similar to the boundary of PPT states. We illustrate this with
some examples on bipartite qubit states, where contours of
constant negativity are plotted on two-dimensional intersections
of the complete state space.
\end{abstract}

\begin{multicols}{2}[]
\narrowtext

In this paper we try to get some insight into the geometrical
structure of entangled states. The main goal will be to
characterize the distance of an entangled state to the set of
separable states. Related questions were addressed in the papers
of Zyczkowski et al.\cite{zy1,zy2,zy3}, Pittenger et al.
\cite{Pittenger} and Witte et al. \cite{Witte} (see also Ozawa
\cite{Ozawa}), although here we attack the problem from a
different perspective.

The concept of negativity will turn out to be very much related
to the Hilbert-Schmidt distance of a state to the set of
separable states. It originates from the observation due to Peres
\cite{Peres} that taking a partial transpose of a density matrix
associated with a separable state is still a valid density matrix
and thus positive (semi)definite. Subsequently
M.Horodecki,P.Horodecki and R.Horodecki \cite{Horodecki} proved
that this was a necessary and sufficient condition for a state to
be separable if the dimension of the Hilbert space does not
exceed $6$. In higher dimensional systems, no easy way of
determining the separability of a state exists due to the
existence of bound entangled states. We will therefore content
ourselves to calculate the Hilbert Shmidt distance of an
entangled state to the set of PPT-states (Remark that in the case
of two qubits no bound entangled states exist).

\begin{figure}
\begin{center}
    \includegraphics[height=5cm]{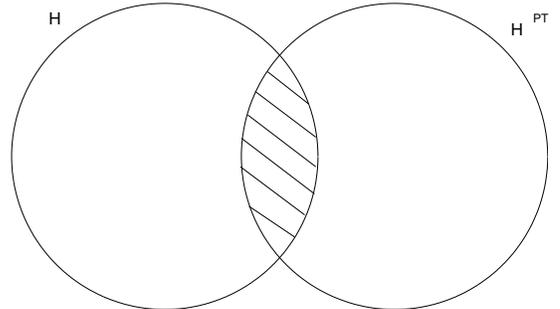}
    \caption{The set of all states is depicted by $H$ and the set of all partial transposed states by $H^{PT}$. The intersection of both is the set of all PPT-states.}
    \label{fig1}
    \end{center}
\end{figure}

This problem is highly related to calculating the distance of an
entangled state to the set of partially transposed states, as the
intersection of the set of all states with the set of all
partially transposed states is equal to the set of all
PPT-states. This is  visualized in figure (\ref{fig1}), where the
boundary of the convex set of states $H$ consists of rank
deficicient states. The set of partially transposed states is
completely isomorf with the set of states, and can be seen as
some kind of reflection of the set of states. The intersection of
both sets is the convex set of PPT-states.

From figure (\ref{fig1}) it is immediatly clear that the distance
of an entangled state to the PPT ones is equal to the distance of
an entangled state to the set of partially transposed states iff
the closest partially transposed state is positive
(semi)-definite; this condition will turn out to be almost always
true.

Let us now calculate the closest partially transposed state to an
entangled state. The key observation is the fact that the Hilbert
Schmidt norm is preserved under the partial transpose map.
Therefore the proposed measure can be defined in the space of
partial transposed density matrices as the minimal Hilbert
Schmidt distance of $\rho^{PT}$ to the surface of positive
(semi)-definite matrices with trace 1, this surface being the
partial transpose of the boundary of PPT-states.

We are therefore looking for the best positive semidefinite
approximation of the indefinite matrix $\rho^{PT}$ in the
Hilbert-Schmidt norm: \be{\min_{\sigma\in
H}\|\rho^{PT}-\sigma\|_2=\sqrt{\tr{(\rho^{PT}-\sigma)^2}}} Writing
the eigenvalue decomposition $\rho^{PT}=UDU^\dagger$, and
absorbing $U$ in $\sigma$, this problem is equivalent to finding
$\sigma$ such that $\|D-\sigma\|$ is minimal. Using the
eigenvalue decomposition $\sigma=VE^2V^\dagger$ with $\tr{E^2}=1$,
this can be written as a Lagrange constrained problem with cost
function:
\be{K=\|D-VE^2V^\dagger\|_2-\lambda\left(\tr{E^2}-1\right)} It is
immediatly clear that the optimal unitary $V$ is given by the
identity: a positive definite matrix remains positive definite if
off-diagonal elements are made zero. Differentiation leads to the
result that the $e_i^2$ are either equal to $0$, either equal to
$d_i+\lambda$. Normalization fixes the value of $\lambda$.
Straightforward calculations show that th $e_j^2$ corresponding
to the negative eigenvalues $d_j$ have to be choosen equal to
zero and the other ones either equal to $d_i+\lambda$ either
equal to $0$, depending on the sign of $d_i+\lambda$.  The
algorithm for finding the closest partially transposed state
therefore becomes:
\begin{enumerate}
\item Calculate the eigenvalue decomposition of $\rho^{PT}=UDU^\dagger$
\item Define $E^2$ as the unique diagonal positive (semi)-definite normalized matrix such that
its elements are $e^2_i=d_i+\lambda$ or $e^2_i=0$.
\item The closest partially transposed state $\rho_s$ is given by
$\rho_s=\left(UE^2U^\dagger\right)^{PT}$. The Hilbert Schmidt
distance between both states is given by
\be{\|\rho-\rho_s\|_2=\sqrt{\frac{(\sum_{i\in I_p}d_i+\sum_{i\in
I_n} d_i)^2}{n_p}+\sum_{i\in I_n} d_i^2},} where $I_n$ is the set
of all indices corresponding to the negative eigenvalues of
$\rho^{PT}$, $I_p$ is the set of indices corresponding to
positive eigenvalues of $\rho^{PT}$ but for which $E^2_i=0$, and
$n_p$ denotes the rank of $E^2$.
\end{enumerate}
If $\rho_s$ is a state, it is guaranteed to be the closest
PPT-state. Numerical investigations show that for example in the
two qubit case the positiveness of $\rho_s$ happens in
approximately 97\% of the cases. If $\rho_s$ is not positive,
then the distance to the set of partially transposed states
calculated is a (fairly good) lower bound on the distance of the
entangled state to the set of PPT-states.

Let us illustrate the above procedure with an example. Say we
want to find the closest biseparable $2\times 4$ state to the
three qubit $W$-state \cite{Dur} $|W\rangle=(|001\rangle
+|010\rangle +|100\rangle)/\sqrt{3}$. The eigenvalue decomposition
of $(|W\rangle\langle W|)^{PT}=UDU^\dagger$, with the partial
transpose operation taken over the 4-dimensional Hilbert space, is
given by:

\bea{D&=&{\rm diag}\small{\ba{cccccccc}{2/3&\sqrt{2}/3&1/3&0&0&0&0&-\sqrt{2}/3}}\\
U&=&\small{\ba{cccccccc}{.&1/\sqrt{2}&.&.&.&.&.&-1/\sqrt{2}\\
1/\sqrt{2}&.&.&1/\sqrt{2}&.&.&.&.\\
1/\sqrt{2}&.&.&-1/\sqrt{2}&.&.&.&.\\
.&.&.&.&1&.&.&.\\
.&.&1&.&.&.&.&.\\
.&1/2&.&.&.&.&1/\sqrt{2}&1/2\\
.&1/2&.&.&.&.&-1/\sqrt{2}&1/2\\
.&.&.&.&.&1&.&.}}} The eigenvalues $E^2$ are readily obtained:
\be{\small{E^2={\rm
diag}\ba{cccccccc}{2/3-\sqrt{2}/9&2\sqrt{2}/9&1/3-\sqrt{2}/9&0&0&0&0&0}}}
Taking the partial transpose leads to the state $\rho_s$, where
we used the notation $c=\sqrt{2}/18$:
\be{\small{\ba{cccccccc}{2c&.&.&.&.&.&.&.\\
.&1/3-c&1/3-c&.&1/9&.&.&.\\
.&1/3-c&1/3-c&.&1/9&.&.&.\\
.&.&.&.&.&.&.&.\\
.&1/9&1/9&.&1/3-2c&.&.&.\\
.&.&.&.&.&c&c&.\\
.&.&.&.&.&c&c&.\\
.&.&.&.&.&.&.&.}}} The eigenvalues of $\rho_s$ are non-negative
and it is possible to show that $\rho_s$ is separable. We have
therefore found the closest biseparable $2\times 4$ state to the
$|W\rangle$-state, and the Hilbert Schmidt distance to it is
equal to $\left(2/3\right)^{3/2}$. Recently, the question arised
whether the set of W-type states is of measure zero. Using the
language of the Hilbert-Schmidt distance, this problem is readily
solved. Indeed, the question is solved if we can prove that the
state obtained by mixing the W-state with a small random
completely separable mixed state remains outside the set of all
convex combinations of biseparable states (with relation to
whatever partition). As there is no biseparable pure state
infinitesimally close to the W-state, and a mixed state not
infinitesimally close to a pure state is always at a finite
distance from whatever pure state, it is proved that the set of
W-type states is indeed not of measure zero. A different proof
was given by Acin et al. \cite{Acinmixed}. Remark that the above
proof is very general and can be used in systems of arbitrary
dimensions: whenever there exists a pure state $\psi_1$ that can
probabilistically be converted into another one $\psi_2$ but not
vice-versa, the set of $\psi_1$-like states minus the set of the
$\psi_2$-like states is of finite measure if there does not exist
a $\psi_2$-like state infinitesimally close to $\psi_1$!

The concept of negativity is also connected to the concept of
robustness of entanglement \cite{Vidalrobust}. Indeed, let us
calculate how much an entangled bipartite state of whatever
dimension has to be mixed with the identity before it gets PPT. In
analogy with the previous derivation of the Hilbert-Schmidt
distance, this amounts to the equivalent problem of how much one
has to mix the partial transpose of $\rho$ with the identity
before it gets positive semi-definite: \be{\min_{t}
(1-t)\rho^{PT}+\frac{t}{4}I_4\geq 0} This problem is readily
solved, and the solution is
\be{t=\frac{|d_{min}|}{|d_{\min}|+\frac{1}{4}}} where $d_{\min}$
is the minimal negative eigenvalue of $\rho^{PT}$. The minimal $t$
is therefore only a function of the negative eigenvalues. A
geometrical implication of this fact is that all surfaces of
constant $d_{\min}$ are similar to the boundary of separable and
entangled states: the set of all states with constant $d_{\min}$
can be generated by extrapolating all lines from the identity to
the boundary of separable states such that the distance of the
extrapolated state to the identity is a constant factor ($>1$) of
the distance of the separable state to the identity.

Let us now move to the case of two qubits. In this case
$\rho^{PT}$ has at most one negative eigenvalue
\cite{Verstraeteconcneg}. Numerical investigations indicate that
in a vast majority of the states the optimal rank of $E^2$ is
equal to three, and if the rank is equal to two it implies that
$\rho_s$ has a negative eigenvalue. For the states for which
$E^2$ is rank 3, it follows that their distance to the set of
partially transposed states is given by
\be{\|\rho-\rho_s\|=\frac{2}{\sqrt{3}}|d_{\min}|} where
$d_{\min}$ is the negative eigenvalue of $\rho^{PT}$. Surfaces of
two-qubit states with constant negativity, defined as
$N=2|d_{\min}|$, have therefore two distinct properties: they are
all similar to each other and the Hilbert-Schmidt distance
between them is almost everywhere constant.

Let us illustrate the above findings by explicitely calculating
some two-dimensional intersections of the set of all bipartite
qubit states including the maximally mixed state. In the
following figures we use the metric based on the Hilbert-Schmidt
distance
$\|\rho_1-\rho_2\|^2=\tr{(\rho_1-\rho_2)^\dagger(\rho_1-\rho_2)}$,
and directions represented orthogonal to each other are orthogonal
in the sense that $\tr{A_1A_2}=0$. Rank deficient density
operators always lie on the boundary of the intersection.

Note that an explicit parameterization of the boundary between the
entangled and separable states can easily be obtained: it is at
most a quartic function of the mixing parameters of the states, as
their analytic expression can be obtained by setting the
determinant of the partial transpose equal to zero.

As a first example we consider the plane containing the maximally
mixed state and the states
\bea{\rho_1&=&\frac{1}{2}\ba{c}{0\\1\\1\\0}\ba{cccc}{0&1&1&0}\nonumber\\
\rho_2&=&\ba{c}{1\\0\\0\\0}\ba{cccc}{1&0&0&0}\label{ff1}}

\begin{figure}
\begin{center}
    \scalebox{.5}{\includegraphics{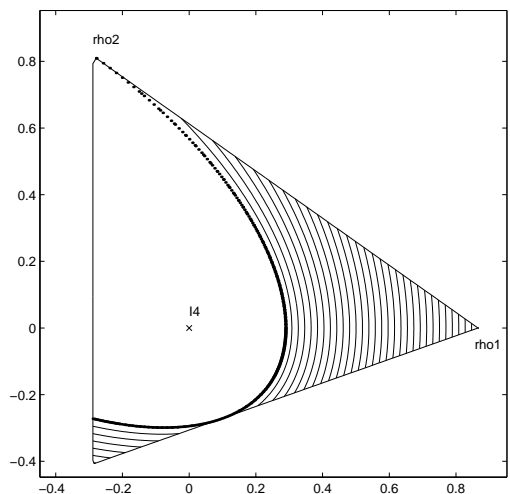}}\\
    \caption{Intersection of the convex set of all states including states (\ref{ff1}) and the maximally mixed state. The contours represent surfaces of constant negativity, the starred line is the boundary between separable and entangled states.}
    \label{f1}
\end{center}
\end{figure}

The plane is plotted in figure (\ref{f1}) and the boundary of all
(rank-deficient) states is given by the solid enveloppe. The
starred line is the boundary between the convex set of separable
states and the convex set of all states. The surfaces of constant
negativity are indeed all similar to this boundary. The fact that
the distance between these surfaces is not constant throughout
the picture indicates that the closest separable states lie in
other planes. Note that the Werner states lie along the line
between the maximally mixed state and the maximally entangled
state $\rho_1$. The thirth extremal point in the undermost left
corner is given by the rank 2 state
\be{\rho=\ba{cccc}{.&.&.&.\\.&\frac{1}{4}&-\frac{1}{4}&.\\.&-\frac{1}{4}&\frac{1}{4}&.\\.&.&.&\frac{1}{2}}}
This state is called a quasi-distillable state and has some
remarkable properties: a single copy of it can be destilled
infinitesimally close to the singlet state
\cite{Horodeckiqd,Verstraete1}, it is the state with minimal
negativity for given entanglement of formation
\cite{Verstraeteconcneg}, and it has furthermore the strange
property that no global unitary operation can increase its
entanglement \cite{Verstraetegenbell}.

\begin{figure}
\begin{center}
    \scalebox{.5}{\includegraphics{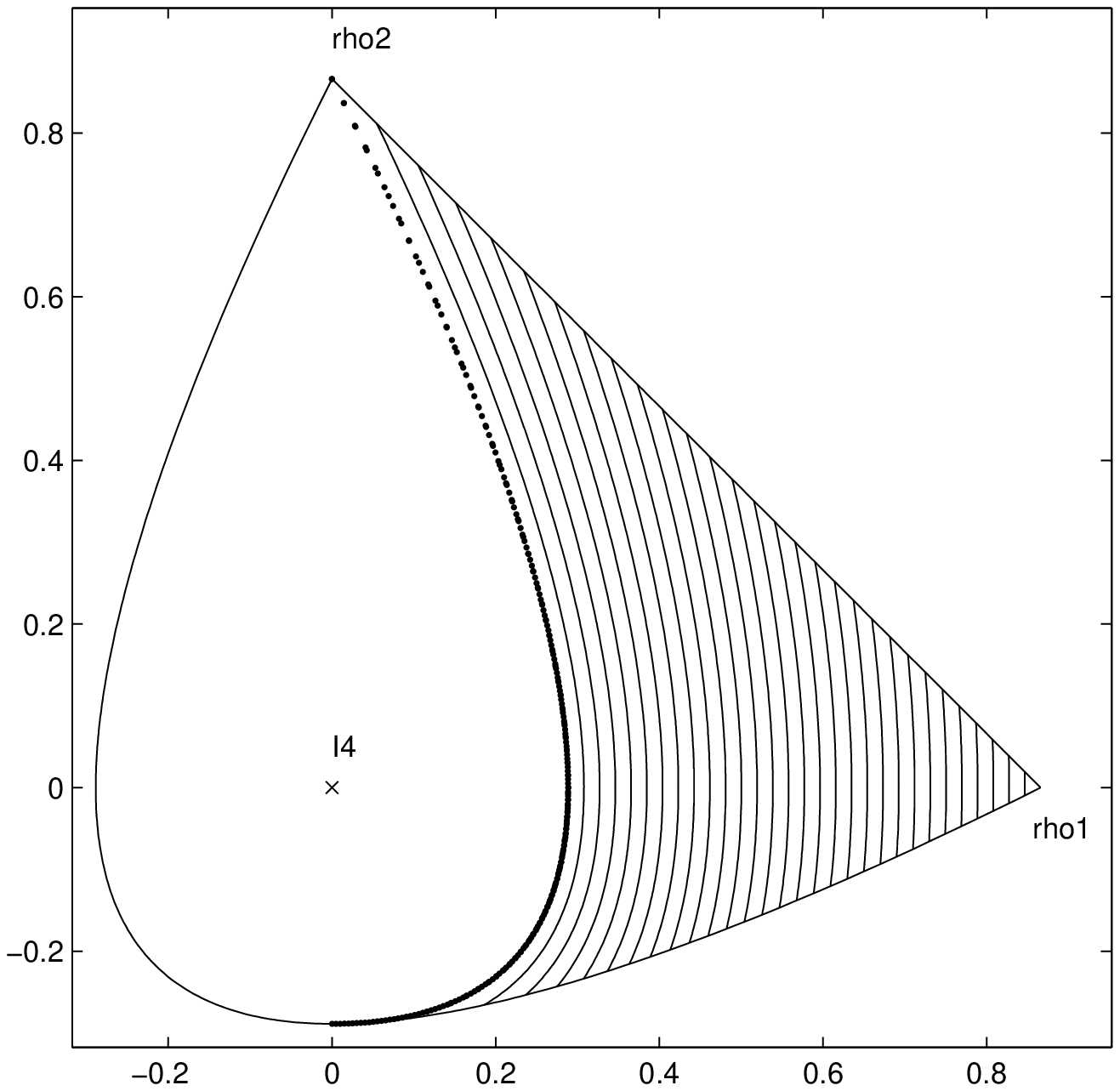}}\\
    \caption{Intersection of the convex set of all states including states (\ref{ff2}) and the maximally mixed state.}
    \label{f2}
\end{center}
\end{figure}

Let us now consider a different plane including the maximally
mixed state and
\bea{\rho_1&=&\frac{1}{2}\ba{c}{0\\1\\1\\0}\ba{cccc}{0&1&1&0}\nonumber\\
\rho_2&=&\frac{1}{2}\ba{c}{1\\1\\0\\0}\ba{cccc}{1&1&0&0}\label{ff2}}
This plane is obtained by rotating the previous plane around the
axis $\rho_1-I_4$. In this case $(\rho_1-I_4)$ is orthogonal to
$(\rho_2-I_4)$, and a completely different picture is obtained as
shown in figure (\ref{f2}).

\begin{figure}
\begin{center}
    \scalebox{.5}{\includegraphics{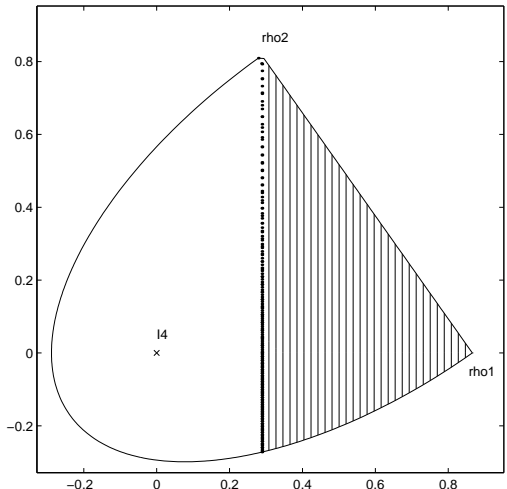}}\\
    \caption{Intersection of the convex set of all states including states (\ref{ff3}) and the maximally mixed state.}
    \label{f3}
\end{center}
\end{figure}

Further rotation of the plane leads to the following states:
\bea{\rho_1&=&\frac{1}{2}\ba{c}{0\\1\\1\\0}\ba{cccc}{0&1&1&0}\nonumber\\
\rho_2&=&\frac{1}{2}\ba{c}{0\\1\\0\\0}\ba{cccc}{0&1&0&0}\label{ff3}}
The intersection of the state space by this plane is shown in
figure (\ref{f3}).

The surfaces of constant negativity become straight lines,
implying that the closest separable states lie in the same plane:
the Hilbert-Schmidt distance betweem the surfaces of constant
negativity has to be constant if they consist of parallel planes.
Using the procedure previously outlined, it is indeed trivial to
check that the separable state closest to the maximally entangled
state $\rho_1$ lies in the defined plane and is given by
\be{\rho_s=\ba{cccc}{\frac{1}{6}&.&.&.\\.&\frac{1}{3}&\frac{1}{6}&.\\.&\frac{1}{6}&\frac{1}{3}&.\\.&.&.&\frac{1}{6}}}

\begin{figure}
\begin{center}
    \scalebox{.5}{\includegraphics{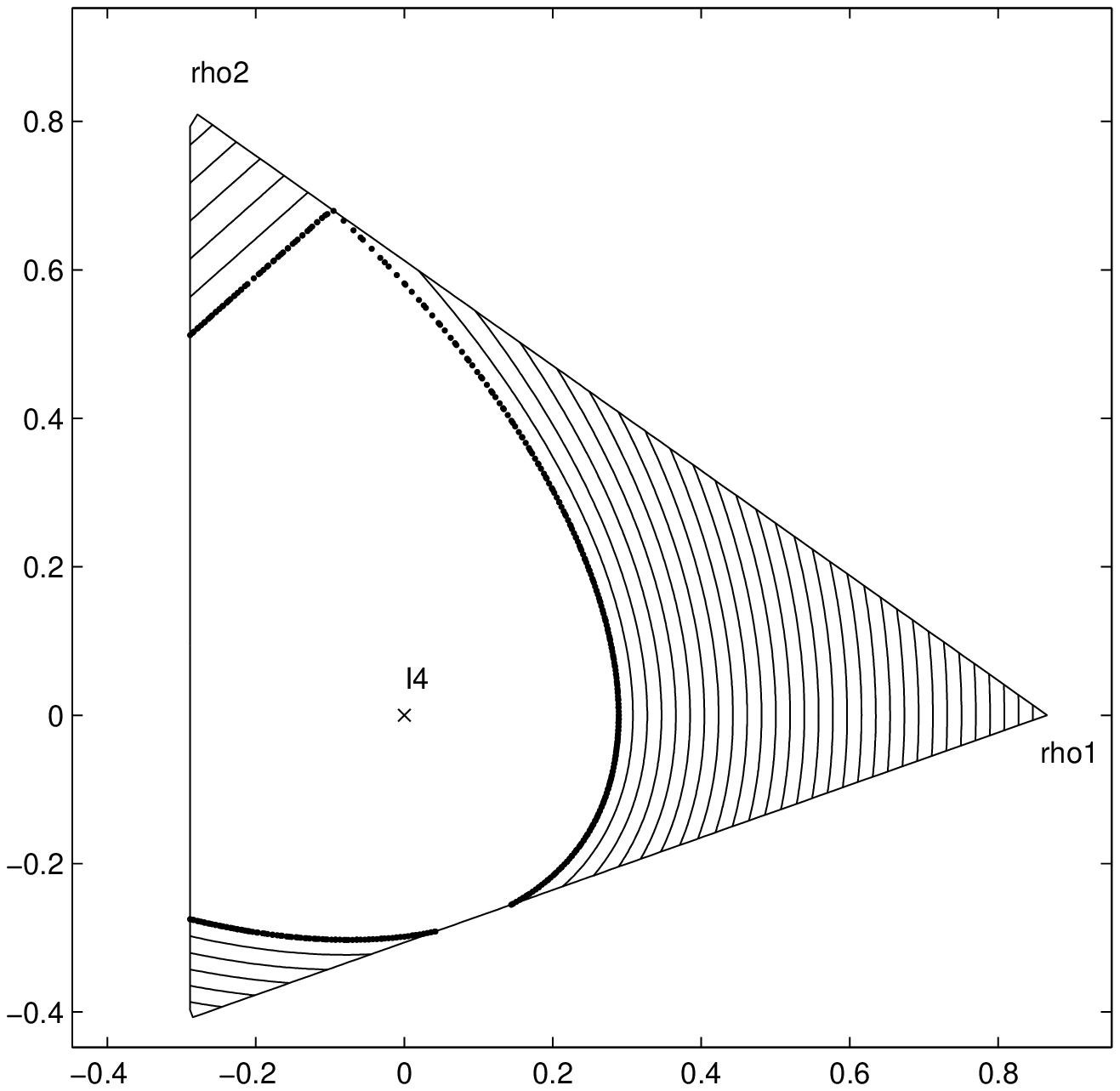}}\\
    \caption{Intersection of the convex set of all states including states (\ref{ff4}) and the maximally mixed state.}
    \label{f4}
\end{center}
\end{figure}

Let us rotate the plane further over the $(\rho_1-I_4)$-axis:
\bea{\rho_1&=&\frac{1}{2}\ba{c}{0\\1\\1\\0}\ba{cccc}{0&1&1&0}\nonumber\\
\rho_2&=&\frac{1}{101}\ba{c}{10\\0\\0\\1}\ba{cccc}{1&0&0&10}\label{ff4}}
The resulting figure (\ref{f4}) combines the features of the
previous figures. Three entangled disconnected regions arise, and
once more we observe the strange shape of the boundary between
entangled and separable states.

\begin{figure}
\begin{center}
    \scalebox{.5}{\includegraphics{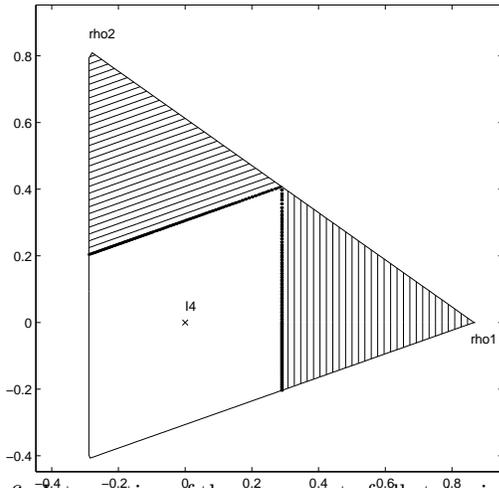}}\\
    \caption{Intersection of the convex set of all states including states (\ref{ff8}) and the maximally mixed state.}
    \label{f8}
\end{center}
\end{figure}

A plane with a highly symmetric contour lines is obtained if
$\rho_1$ and $\rho_2$ are both taken to be maximally entangled
states:
\bea{\rho_1&=&\frac{1}{2}\ba{c}{0\\1\\1\\0}\ba{cccc}{0&1&1&0}\nonumber\\
\rho_2&=&\frac{1}{2}\ba{c}{0\\1\\-1\\0}\ba{cccc}{0&1&-1&0}\label{ff8}}
Indeed, only straight lines are obtained in figure (\ref{f8}).
The thirth extremal state is in this case given by $\rho={\rm
diag}[1/2;0;0;1/2]$.

\begin{figure}
\begin{center}
    \scalebox{.5}{\includegraphics{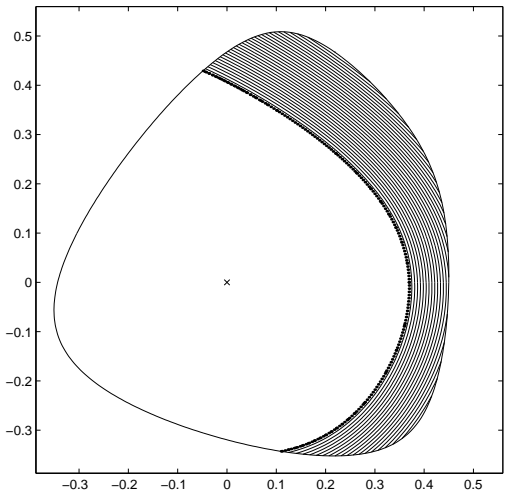}}\\
    \scalebox{.5}{\includegraphics{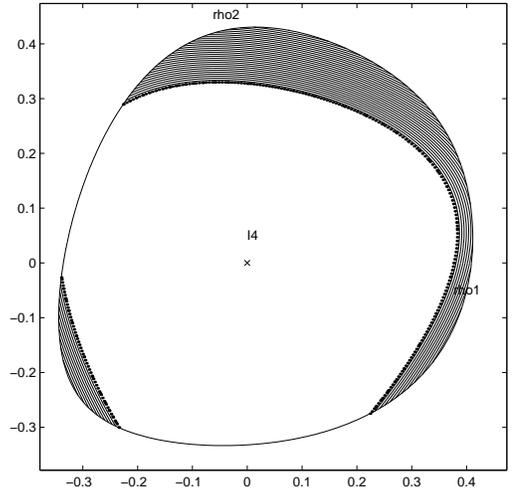}}\\
    \caption{Contour plots of the negativity on random planes including the maximally mixed state.}
    \label{f9}
\end{center}
\end{figure}

At last, we choose two random planes through the maximally mixed
state and plot them in figure (\ref{f9}).
The similarity of all planes with constant negativity is clearly
illustrated.

We acknowledge interesting discussions with K. Audenaert, A.
Pittenger and K. Zyczkowski, who gave the idea of considering 2-D
intersections of the state space.

\end{multicols}
\end{document}